\documentclass[dvips,12pt] {article}
\usepackage{epsfig}
\usepackage{rotate}
\usepackage{a4}
\def\lsim{\mathrel{\rlap{\raise 2.5pt \hbox{$<$}}\lower 2.5pt
\hbox{$\sim$}}}
\renewcommand{\thefootnote}{\fnsymbol{footnote}}
\newcommand{\eV}{\mbox{eV}}

\def\air{\vphantom{\bigg|}}

\begin{document}
\thispagestyle{empty}

\begin{flushright}
CERN-TH/98-200 \\[2mm]
hep-ph/9806339 \\
\end{flushright}
 \vspace*{3cm}
 \begin{center}
 {\bf \Large
 Solar-neutrino oscillations \\[5mm] and third-flavour admixture}
 \end{center}
 \vspace{1cm}
 \begin{center}
 P.~Osland$^{a,b}$ \ \
 \ and \
 G. Vigdel$^{a}$ \\
 \vspace{1cm}
$^{a}${\em
 Department of Physics, University of Bergen, \\
      All\'{e}gaten 55, N-5007 Bergen, Norway}
\\
\vspace{.3cm}
$^{b}${\em
Theoretical Physics Division, CERN, 
CH-1211 Geneva 23, Switzerland }
\end{center}
\hspace{3in}

\begin{abstract}
With one $\Delta m^2$ of the appropriate order of
magnitude to solve the atmospheric neutrino problem, we study the
resulting three-generation vacuum-oscillation fit to the solar neutrino 
flux.
An explanation of the atmospheric neutrino composition
in terms of pure $\nu_\mu \rightarrow \nu_\tau$ oscillations
is easily compatible with the well-known 
two-flavour oscillation solution for solar neutrinos.  
The allowed parameter region in the 
$\sin^2(2\theta_{12})$--$\Delta m^2$ plane
changes little with increasing values of the mixing element $U_{e3}$, 
provided this is less than about 0.4.
We find that the threefold maximal mixing is disfavoured.
\end{abstract}
\vfil

\begin{flushleft}
CERN-TH/98-200 \\[2mm]
June 1998          \\
\end{flushleft}

\renewcommand{\thefootnote}{\arabic{footnote}}
\section{Introduction}
While neutrino-oscillation experiments tend to suffer from
systematic errors that are difficult to estimate,
there are at least two distinct neutrino problems that
deserve serious consideration:
the deficit of solar neutrinos 
and the atmospheric neutrino composition.
The former has been around for many years.
It started with the pioneering Homestake detector
and later gained significance with the water-scattering
and gallium experiments \cite{solar}.

The latter anomaly is an inconsistency between the number of detected 
neutrinos of the muon family and those of the electron family,
assumed to be generated by cosmic-ray hadrons 
impinging on the atmosphere. The ratio of ratios
\begin{equation}
R = \frac{(N_\mu/N_e)_{\rm observed}}{(N_\mu/N_e)_{\rm predicted}},
\end{equation} 
where $N_\mu$ is the number of $\mu$-like events and $N_e$ the
number of $e$-like events, should be, according to theory, close to 1, 
but is instead around $\sim 0.6$ \cite{ref:super-kamiokande}.

If one analyses the solar-neutrino deficit 
(a summary of the chlorine \cite{ref:homestake},
gallium \cite{ref:gallex,ref:sage} and water-scattering 
experiments \cite{ref:kamiokande,ref:SK-sun}
is given in Table \ref{experiment}) 
in terms of two-flavour mixing, the vacuum-oscillation
solution leads to a set of disconnected regions
in the two-dimensional parameter space spanned by $\sin^2(2\theta)$
and $\Delta m^2$, according to the $\nu_e$ survival probability
\begin{equation} \label{protwogen}
 P_{\nu_e\rightarrow \nu_e}(t) 
= 1 - \sin^2(2\theta)\sin^2\left(\frac{\Delta m^2 t}{4E}\right),
\end{equation}
where $\theta$ is the mixing angle, $\Delta m^2$ the difference of the
squared masses, $t$ the time since creation and $E$ the neutrino energy.
The favoured region is $\sin^2(2\theta)\simeq0.8$--0.9,
with $\Delta m^2\simeq6\times10^{-11}~\eV^2$, but neighbouring
regions that differ from $\Delta m^2$ by 
$n\times\delta m^2$ can yield comparable
values of $\chi^2$, provided $n$ is a small integer, and
$\delta m^2$ is of the order of $1.4\times10^{-11}~\eV^2$
(cf.\ Fig.~1a)\footnote{The MSW \cite{MSW} 
interpretation of the data leads to values of $\Delta m^2$
that are several orders of magnitude higher.}.
The scale $\delta m^2$ is given by the characteristics of the detectors, 
the neutrino energy (of the order of a MeV)
and the distance $R$ from the Sun to the Earth:
$\delta m^2\simeq 4\pi E\hbar c/R$.

The probability expression (\ref{protwogen}) and Fig.~1a show the 
`just so' character of vacuum oscillations, the mass-squared difference 
has to be fine-tuned in relation to the Sun--Earth distance if this model 
shall solve the solar-neutrino problem \cite{just-so}.
(The allowed region in the upper part of Fig.~1a, 
where $\Delta m^2>2\times 10^{-10}~\eV^2$, is a new feature of the
most recent data \cite{Bahcall}.)

With three flavours of neutrinos, one may consider the general
three-flavour mixing, in which a state, which is of pure flavour
$\alpha$ ($\alpha=e$, $\mu$, $\tau$) at $t=0$, evolves as
\begin{equation}\label{mixing}
  \nu(t) = \sum_j U_{\alpha j}e^{-iE_jt}\nu_j,
\end{equation}
where the summation index $j$ runs over mass states.
We write the unitary mixing matrix as
\begin{equation} \label{Ugen}
U=\left[ 
\begin{array}{ccc} 
U_{e1} & U_{e2} & U_{e3} \\ 
U_{\mu 1} & U_{\mu 2} & U_{\mu 3} \\ 
U_{\tau 1} & U_{\tau 2} & U_{\tau 3} 
\end{array}\right].
\end{equation}
The full parameter space then consists of three angles (we ignore
$CP$-violating effects)
and two independent squared-mass differences (the energy can be 
approximated as $E_j\simeq p+m_j^2/2E$). In particular,
the survival probability of an electron neutrino takes the form
\begin{eqnarray} \label{probunav} 
P_{\nu_e\rightarrow \nu_e}(t)
&=& 1
-4\biggl[U_{e1}^2U_{e2}^2\sin^2\left(\frac{\Delta m^2_{21}t}{4E}\right)
+U^2_{e1}U^2_{e3}\sin^2\left(\frac{\Delta m^2_{31}t}{4E}\right)
\nonumber \\ 
&&
\phantom{1-4}
+U_{e2}^2U_{e3}^2\sin^2\left(\frac{\Delta m^2_{32}t}{4E}\right)
\biggr],
\end{eqnarray}
where $\Delta m_{ij}^2 = m_{i}^2 - m_{j}^2$.
Unitarity and data on other phenomena, in particular
on the atmospheric neutrino flux, lead to constraints on these
parameters.

We denote the mass states $\nu_1$, $\nu_2$ and $\nu_3$, 
according to their intrinsic mass ratio:
their respective mass eigenvalues will be sorted as
$m_1 < m_2 < m_3$.
For the mixing matrix we use the
form advocated for quarks by the Particle Data Group 
\cite{PDG}\footnote{There are
nine structurally different ways to express the mixing matrix in terms of
mixing angles.  A discussion of the parametrization of 
flavour mixing in the quark sector is presented in \cite{Fritzsch},
where also $CP$ violation is included.}:
\begin{eqnarray}
\label{U}
U&=&U_{\mu\tau}(\theta_{23})U_{e\tau}(\theta_{13})U_{e\mu}(\theta_{12}) 
\nonumber  \\
&=& \left[
\begin {array}{ccc} 
c_{12}\, c_{13} & s_{12}\,c_{13} & s_{13}\\
-c_{12}\,s_{13}\,s_{23}-s_{12} \,c_{23} &
-s_{12}\,s_{13}\,s_{23}+c_{12}\,c_{23} & c_{13}\,s_{23}\\
-c_{12}\,s_{13}\,c_{23}+s_{12}\,s_{23}&
-s_{12}\,s_{13}\,c_{23}-c_{12}\,s_{23}& c_{13}\,c_{23}
\end{array}\right],
\end{eqnarray}
where $c_{12}= \cos\theta_{12}$, $c_{13}= \cos\theta_{13}$, etc., and
the mixing angles $\theta_{12},\theta_{13} \mbox{ and } \theta_{23}$ 
are in the interval $[0,\pi/2]$. 
\section{Three-parameter accommodation of the solar neutrino flux}
A three-flavour analysis of the solar-neutrino oscillations will
restrict the allowed values of the elements in the mixing matrix.  
Because of unitarity, the solar electron-neutrino survival
probability can always be expressed in terms of only {\it two}
mixing angles, here taken to be $\theta_{12}$ and $\theta_{13}$.
We shall assume that the dominant mixing 
(represented in Fig.~1a) is a mixing between the first two families, 
and ask 
(i) `How will the goodness of fit depend on $U_{e3}$' and 
(ii) `How will the allowed regions of Fig.~1a change as one 
increases the strength of this mixing?'

Before we address these questions, we want to introduce some further
restrictions on the neutrino parameters:
we reserve one $\Delta m^2$
for the vacuum-oscillation solution of the solar-neutrino problem,
$\Delta m_{21}^2 \sim 10^{-11}~\mbox{eV}^2$.
The suppression of atmospheric muon neutrinos relative to electron
neutrinos suggests that also $\nu_\mu$ oscillates into some other
flavour of neutrinos, but (since the ratio $t/E$ relevant 
to atmospheric neutrinos is orders of magnitudes smaller than that
for solar neutrinos)
with a mass-squared difference that is
many orders of magnitude larger than is required to explain
the solar-neutrino deficit.
It is therefore natural to assume that other squared-mass differences
are much larger:
\begin{equation}\label{assumption}
\Delta m^2_{21} \sim 10^{-11}~\mbox{eV}^2 
\ll \Delta m^2_{32}\lsim \Delta m^2_{31}, 
\end{equation} 
such that when Eq.~(\ref{probunav}) is applied to the case of
solar neutrinos, the last two terms can be
represented by their average values:
\begin{equation}
\label{avsurvival} 
P^{\rm sun}_{\nu_e \rightarrow \nu_e}(t) 
=1-\sin^2(2\theta_{12})\cos^4\theta_{13}
\sin^2\left(\frac{\Delta m_{21}^2 t}{4E}\right) 
- \frac{1}{2}\sin^2(2\theta_{13}).
\end{equation}
Here, the last term represents a non-zero component of
the $\nu_3$ mass state in the electron-neutrino state, 
and vanishes with $|U_{e3}|^2$.
For $\theta_{13}=0$ we get the familiar two-family
mixing of Eq.~(\ref{protwogen}).

While a simultaneous fit to both solar- and atmospheric-neutrino
fluxes also would involve the angle $\theta_{23}$ as well as 
two mass-squared differences, under the assumption (\ref{assumption})
the above formula (\ref{avsurvival})
provides a three-parameter accommodation of the solar flux.

\begin{figure}[t]
\begin{center}
\begin{picture}(0,0)(14,29)
  \put(1.5,0.7){\mbox{\epsfysize=35cm\epsffile{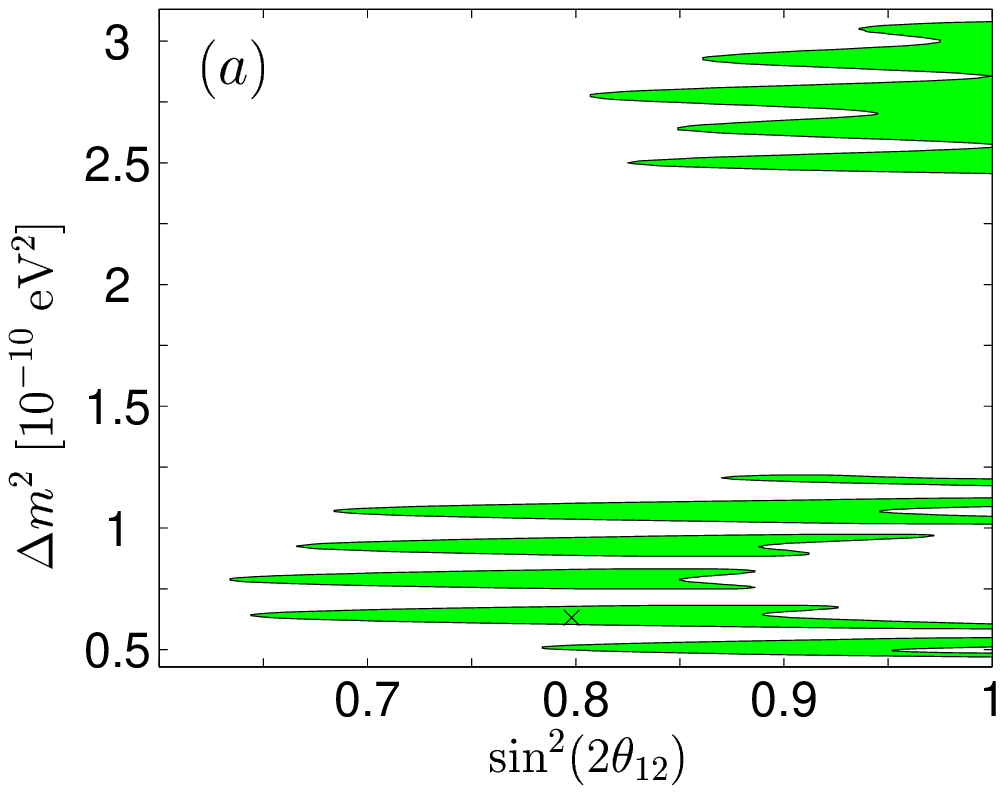}}}
  \put(1,-9.5){\mbox{\epsfysize=35cm\epsffile{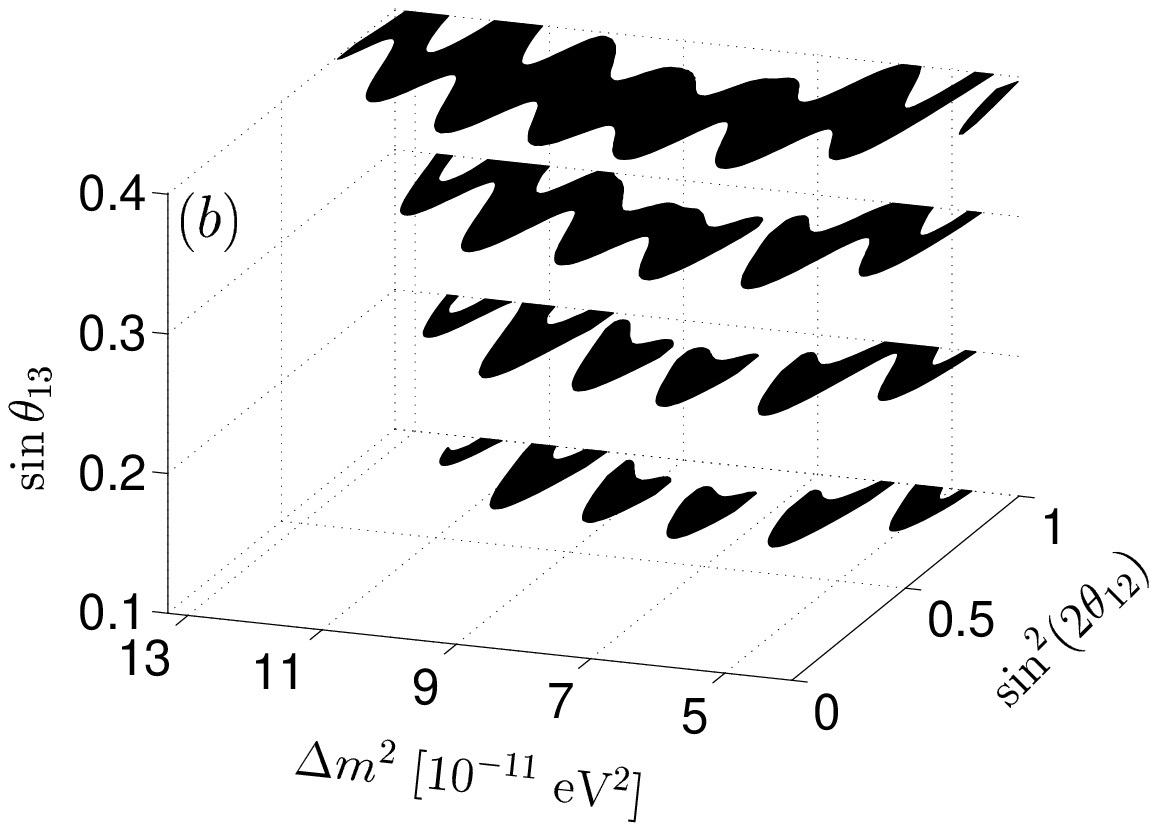}}}
\end{picture}
\end{center}
\vspace*{18cm}
\caption{Allowed parameter space within 95\% C.L.
The upper figure illustrates the two-generation case, 
the cross marks the position of $\chi_{\rm min}^2$. 
The lower figure shows (under the assumption
leading to Eq. (\ref{avsurvival})) the dependence
on the mixing element $U_{e3}=\sin\theta_{13}$, which increases in
steps of 0.1 from 0.1 to 0.4.
The regions merge as the third mass state $\nu_3$ gets more involved
in the mixing.}
\label{parameterspace1}
\end{figure}

We show in Fig.~\ref{parameterspace1}b the results of two-parameter 
fits to the solar-neutrino data. 
(The high region in $\Delta m^2$, cf.\ Fig.~\ref{parameterspace1}a,
is not displayed here.)
In our calculations we use the solar-neutrino flux
as given by the 1998 Standard Solar Model (SSM) 
by Bahcall, Basu and Pinsonneault \cite{Bahcall}, 
and for the detection rates the values given in Table \ref{experiment}.
The second mixing angle, $\theta_{13}$, 
is held fixed at different values, and the fit is performed
in the variables $\sin^2(2\theta_{12})$ and $\Delta m^2_{21}$.
It is seen that regions that are separated in the two-family case,
can be smoothed out to one big region when the more general case of 
three flavours is considered. 
However, there is a slight worsening of the quality
of the fit as the coupling to the third state is increased.
For $\sin\theta_{13}=0.0$, 
0.2, and 0.4, the values of $\chi_{\rm min}^2$ are 3.6,
4.2 and 5.6, respectively.
For two degrees of freedom, 4 experiments (SAGE and GALLEX are combined) 
minus 2 parameters, the probability of getting $\chi^2$ larger than 
those values by chance, are 17, 12 and 6\%, respectively.  
None of these is particularly good, but they are not excluded.

Since there is some uncertainty associated with the important boron flux
\cite{Bahcall},
we have checked how the fit changes under 10\% rescalings of
the $^8{\rm B}$ flux.
In Table~2 we show the resulting $\chi_{\rm min}^2$ and 
goodness of fit for three different fractions of the SSM boron flux 
(keeping the other components of the flux fixed),
for two different values of $U_{e3}$.  For the lowest of these fractions, 
the goodness of fit is appreciably better than for the standard 
$^8{\rm B}$-flux.
It is also seen that, for the lowest of our tabulated boron flux values,
the fit deteriorates less when $\sin\theta_{13}$ is non-zero.

A different representation, for several values of $\sin\theta_{13}$,
is given in Fig.~\ref{fig2}, including also the higher-$\Delta m^2$
region.
As $\sin\theta_{13}$ increases, the allowed regions get broader,
but also the fit gets poorer.

It has been argued that the data suggest `maximal mixing' \cite{max-mix}.
In our variables, maximal mixing corresponds to $\sin^2(2\theta_{12})=1$
and $\sin\theta_{13}=1/\sqrt{3}\simeq0.58$.
We find that the fit gets rather poor for such parameters,
with $\chi_{\rm min}^2= 11$.

\section{Long-baseline experiments}
The future long-baseline experiments \cite{LB} 
will provide new restrictions on these mixing elements.
For example, if we assume that a long-baseline experiment
is such that oscillations other than those involving 
\begin{equation}
\Delta M^2\equiv\Delta m^2_{31}\simeq\Delta m^2_{32}
\end{equation}
can be neglected,
then the condition (\ref{assumption}) leads to the following transition
probabilities:
\begin{equation}\label{mutoe}
P^{\rm L.B.}_{\nu_\mu \rightarrow \nu_e}(t) 
=\sin^2\theta_{23}\sin^2(2\theta_{13})
\sin^2\left(\frac{\Delta M^2t}{4E}\right),
\end{equation}
\begin{equation}\label{mutotau}
P^{\rm L.B.}_{\nu_\mu \rightarrow \nu_\tau}(t) 
=\cos^4\theta_{13}\sin^2(2\theta_{23})
\sin^2\left(\frac{\Delta M^2t}{4E}\right),
\end{equation}
and the survival probability:
\begin{equation}\label{etoe}
P^{\rm L.B.}_{\nu_e \rightarrow \nu_e}(t)
=1-\sin^2(2\theta_{13})
\sin^2\left(\frac{\Delta M^2t}{4E}\right).
\end{equation}
Note that the coefficient in Eq.~(\ref{mutoe})
can be written as $4|U_{e3}|^2|U_{\mu3}|^2$.
The $\nu_\mu\to\nu_e$ transition can thus be given a simple
physical interpretation, valid for small values of
the mixing angle $\theta_{13}$.
When $\Delta m^2_{21}t/E \ll 1$, the transition can be interpreted 
as an `indirect' transition,
involving the admixtures of $\nu_3$ as given by
$\theta_{13}$ and $\theta_{23}$.
This is different from the case of $\Delta m^2_{21}t/E \simeq 1$, 
where the `direct' transition $\nu_e\to\nu_\mu$ is effective
via the states $\nu_1$ and $\nu_2$.

The atmospheric-neutrino data are usually explained in terms of
$\nu_\mu \rightarrow \nu_e$ or $\nu_\mu \rightarrow \nu_\tau$
oscillations with a large amplitude \cite{yasuda,teshima}.
When applying Eq.~(\ref{avsurvival}) to explain the solar-neutrino deficit,
the transition $\nu_\mu \rightarrow \nu_\tau$ is preferable for the 
atmospheric neutrinos, because it allows for a small 
or vanishing value of $\theta_{13}$, which gives the best fit of the
observed solar neutrino rate.
An early three-flavour analysis (not including data from
Super-Kamiokande) gave for the best fit a value for $|U_{e3}|^2$
(or $\sin^2\theta_{13}$) in the range 0.06--0.14 \cite{bilenky}.
More recent studies (including Super-Kamiokande data) prefer
values for $\sin\theta_{13}$ of 0.1 \cite{yasuda} and 
$<0.1$ \cite{teshima}.
As we have seen, these low values of $\sin\theta_{13}$
also lead to acceptable fits to the solar-neutrino data,
with $\chi_{\rm min}^2=3.6$--5.6.

\begin{figure}[t]
\begin{center}
\begin{picture}(0,0)
\put(-11,-24)
{\mbox{\epsfysize=30cm\epsffile{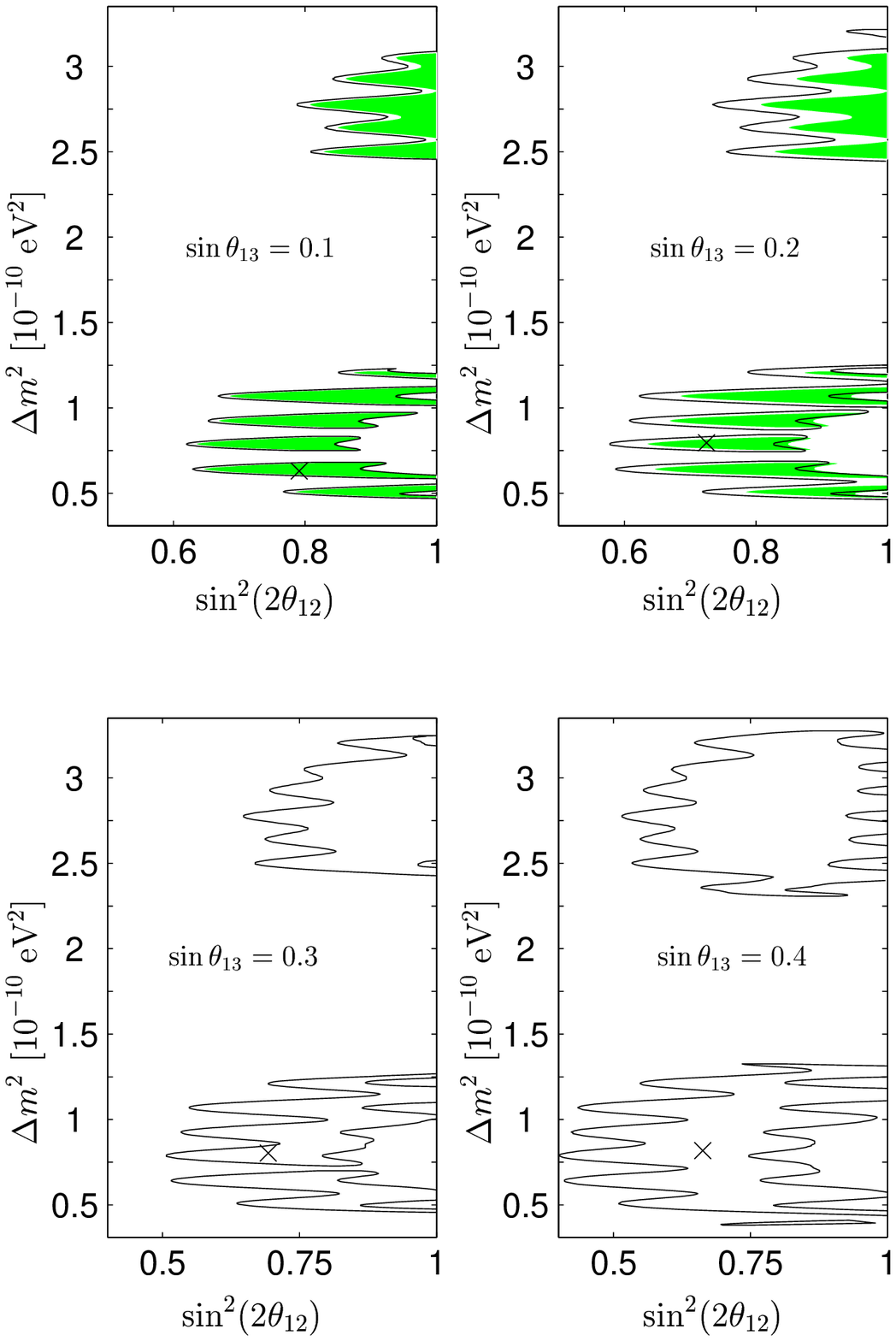}}}
\end{picture}
\end{center}
\vspace*{18.8cm}
\caption{Allowed parameter space within 95\% C.L.\ for $\Delta M^2
\gg \Delta m^2$,  for different
values of $U_{e3}$ ($=\sin\theta_{13}$).
In the two upper boxes,
the allowed regions for the two-generation oscillation 
(Fig.\ \ref{parameterspace1}a) are also shown, as shaded areas.
Note different horizontal scales. The crosses show the 
best-fit point.}
\label{fig2}
\end{figure}

In the long-baseline experiments \cite{LB,ref:CHOOZ}, 
neutrinos from either an accelerator or a reactor will be counted
after travelling a `long' distance, from 1 km (for CHOOZ)
to 730 km (for the MINOS experiment).
In fact, the recent data from the CHOOZ experiment, where the survival 
probability $\bar{\nu}_e \rightarrow \bar{\nu}_e$ is measured,
does not exclude any of the cases studied in Fig.~\ref{fig2}, provided 
$\Delta M^2 < 9\times 10^{-4}~\mbox{eV}^2$.  
Indeed, no evidence for oscillations was found for 
$\Delta M^2 > 2 \times 10^{-2}~\mbox{eV}^2$ {\it and} 
$\sin^2(2\theta_{13}) > 0.18$.
This means that $\sin\theta_{13}$ has to be either small ($<0.22$)
or quite big ($>0.98$) (or $\Delta M^2 < 2 \times 10^{-2}~\mbox{eV}^2$).
With the large-angle solution, Eq~(\ref{avsurvival}) gives a poor
fit to the solar-neutrino data, this alternative should
therefore be disregarded.  The remaining low-angle solution is also
preferred in the analysis of the atmospheric-neutrino data \cite{yasuda}.
In this case (with $\Delta M^2 > 2 \times 10^{-2}~\mbox{eV}^2$) the lower
two boxes of Fig.~\ref{fig2} would be excluded.
The Bugey reactor experiment \cite{bugey}, which is less sensitive,
excludes $\Delta M^2$ down to about $3\times 10^{-2}~\mbox{eV}^2$
for $\sin\theta_{13} > 0.16$, consistent with CHOOZ.

A measurement of a non-zero value for $P_{\nu_\mu \rightarrow \nu_e}$
will imply a non-zero value for $\theta_{13}$.
Such information, which would be a major result from the planned
long-baseline experiments,
combined with Eq.~(\ref{avsurvival}), can help us to further restrict 
the allowed parameter space relevant to the solar neutrinos.

We note that, for $\nu_e\leftrightarrow\nu_\mu$ oscillations, the K2K
experiment \cite{K2K} will have a sensitivity of $\sin^2(2\theta)>0.1$ 
and the MINOS experiment could probe values as low as 
$\sin^2(2\theta) = 0.01$.
In our parametrization, this means that they would be sensitive
to values of $\sin\theta_{13}$ as low as 0.16 and 0.05.
If this mixing element should turn out to be significantly
larger, the vacuum-oscillation solution
to the solar-neutrino problem would have serious difficulties.

\section{Concluding remarks}
The data from the solar-neutrino detectors can very well be reconciled 
with the atmospheric-neutrino data.  
If we assume that the solution of the latter anomaly is 
the $\nu_\mu \rightarrow \nu_\tau$ transition (which is induced by
the direct mixing $\theta_{23}$ and gives
the best fit), then the vacuum-oscillation solution for the
atmospheric and the solar-neutrino problems can be regarded as 
two separate two-generation oscillations.  
Even if the transition $\nu_\mu \rightarrow \nu_e$, which for
$\Delta m^2_{21}L/E\ll 1$ is induced by the two-step process
of $\nu_e\nu_\tau$ and $\nu_\mu\nu_\tau$ mixing (with angles $\theta_{13}$ 
and $\theta_{23}$, respectively) should turn out 
to take place, the rate has to be quite large before it can 
severely alter the fit of the vacuum-oscillation solution to the
solar-neutrino problem. 

When the $\nu_e\nu_\tau$ mixing angle $\theta_{13}$ is increased from zero,
the allowed ranges of $\Delta m^2_{21}$ merge to a wider one,
but beyond $\sin\theta_{13}\simeq0.5$, which is well within
the sensitivity that will be reached by the K2K experiment, 
the fit deteriorates rapidly.

\newpage

\clearpage
\begin{table}[t]
\begin{center}
\begin{tabular}{|l|l|l|l|l|} \hline
Experiment &  Counting rate & SSM \cite{Bahcall} & Counted/SSM \\ 
\hline\hline 
\air
Homestake \cite{ref:homestake} &
$2.54\pm 0.14 \pm 0.14$ & $7.7 ^{+1.2}_{-1.0}$ & $0.330\pm 0.026$ \\
\hline 
\air
GALLEX \cite{ref:gallex} &
$69.7 \pm 6.7 ^{+3.9}_{-4.5}$ & $129 ^{+8}_{-6}$ & 
$0.540 ^{+0.060}_{-0.063}$ \\
\hline 
\air
SAGE \cite{ref:sage} &
$72^{+12}_{-10}$ $\!\!^{+5}_{-7}$ & $129 ^{+8}_{-6}$ & $0.558
^{+0.100}_{-0.095}$\\ 
\hline 
\air
Kamiokande \cite{ref:kamiokande} &
$2.80 \pm 0.19\pm 0.33$ & $5.15 ^{+0.98}_{-0.72}$ & $0.544 \pm 0.074$\\
\hline
\air
Super-Kamiokande \cite{ref:SK-sun} &
$2.42 \pm 0.06 ^{+0.10}_{-0.07}$ & $5.15 ^{+0.98}_{-0.72}$ &
$0.470 ^{+0.023}_{-0.017}$\\ 
\hline
\end{tabular}
\caption{Solar neutrino fluxes.
Measured and SSM-predicted event rates
are for the radio-chemical detectors measured in units of SNU 
(1 event per second per $10^{36}$ target atoms). 
For the scattering detectors (the last two entries)
the fluxes are in units of $10^6 \mbox{ cm}^{-2}{\rm\ s}^{-1}$.  
In the ratio of the observed to the predicted values, 
only experimental uncertainties are included.}
\label{experiment}
\end{center}
\end{table}
\vspace*{20mm}

\begin{table}[h]
\begin{center}
\begin{tabular}{l|ccc|ccc|} 
\cline{2-7}
 &\vspace*{-4.5mm} & & & & & \\
\cline{2-7}
& \multicolumn{3}{c|}{$\sin\theta_{13}=0.0$} &
\multicolumn{3}{c|}{$\sin\theta_{13}=0.3$} \\ 
\cline{2-7} \cline{2-7} 
& $f_{\rm B}=0.9$ &$f_{\rm B}=1.0$ & $f_{\rm B}=1.1$ & 
$f_{\rm B}=0.9$ & $f_{\rm B}=1.0$& $f_{\rm B}=1.1$ \\ \cline{1-2}
\cline{2-7}
\multicolumn{1}{|c|}{$\chi^2_{\rm min}$} & 2.0 & 3.6& 5.8 & 2.9& 4.8& 7.2 \\
\multicolumn{1}{|c|}{$P(\%)$} & 38 & 17 & 5.5  & 23 & 9   & 3 \\ \hline\hline
\end{tabular}
\caption{Best fits for different fractions, $f_{\rm B}$, 
of the SSM $^8{\rm B}$ flux,
for two different values of $\sin\theta_{13}=U_{e3}$.  
The lowest row gives the goodness of fit.}
\end{center}
\label{GOF} 
\end{table}

\newpage
\clearpage



\begin{thebibliography}{99}

\bibitem{solar}
{\it Solar Neutrinos: The First 30 Years},
edited by: J. N. Bahcall et al.\ (Addison Wesley, New York, 1994); \\
J. N. Bahcall, Phys.\ Lett.\ {\bf B 338}, 276 (1994).

\bibitem{ref:super-kamiokande}
Y. Fukuda et al.\ (Super-Kamiokande Collaboration),
ICRR-411-98-7 (1998),
hep-ex/9803006 and hep-ex/9805006.

\bibitem{ref:homestake} 
B. T. Cleveland, Nucl.\ Phys.\ (Proc.\ Suppl.) {\bf B 38}, 47 (1995);\\
K. Lande et al., in {\em Neutrino '96}, 
Proceedings of the 17th International Conference on
Neutrino Physics and Astrophysics, Helsinki, 1996, edited by
K. Huitu, K. Enqvist and J. Maalampi (World Scientific, Singapore,
1997).

\bibitem{ref:gallex} 
P. Anselmann et al., Phys.\ Lett.\ {\bf B 342}, 440 (1995); \\
W. Hampel et al., Phys.\ Lett.\ {\bf B 388}, 384 (1996).

\bibitem{ref:sage} 
J. N. Abdurashitov et al., Phys.\ Lett.\ {\bf B 328}, 234 (1994).
V. Gavrin et al., in {\em Neutrino '96}, {\em op.\ cit.}

\bibitem{ref:kamiokande} 
Y. Fukuda et al., Phys.\ Rev.\ Lett.\ {\bf 77}, 1683 (1996).

\bibitem{ref:SK-sun}
Y. Fukuda et al., 
hep-ex/9805021, submitted to Phys.\ Rev.\  Lett.
See also \\
Y. Itow (contains S.K.\ data until 20 October 1997)\\
http://www-sk.icrr.u-tokyo.ac.jp/doc/sk/pub/pub\underline{\hspace{2mm}}sk.html

\bibitem{MSW}
S. P. Mikheyev and A. Yu.\ Smirnov, Yad.\ Fiz.\ {\bf 42}, 1441 (1985)
[Sov.\ J. Nucl.\ Phys.\ {\bf 42}, 913 (1985)],
Nuovo Cimento {\bf 9C}, 17 (1986); \\
L. Wolfenstein, Phys.\ Rev.\ {\bf D 17}, 2369 (1978),
{\it ibid.} {\bf 20}, 2634 (1979).

\bibitem{just-so}
V. Barger, R. J. N. Phillips and K. Whisnant,
Phys.\ Rev.\ {\bf D 24}, 538 (1981);
S. L. Glashow and L. M. Krauss, 
Phys.\ Lett.\ {\bf 190 B}, 199 (1987).

\bibitem{Bahcall} 
J. N. Bahcall, S. Basu and M. H. Pinsonneault, astro-ph/9805135.

\bibitem{PDG} 
Particle Data Group (R. M. Barnett et al.),  
Phys.\ Rev.\ {\bf D 54}, 1 (1996).
 
\bibitem{Fritzsch} 
H. Fritzsch and Z.-z.\ Xing, 
Phys.\ Rev.\ {\bf D 57}, 594 (1998); 
A. Rasin, hep-ph/9708216.

\bibitem{max-mix}
P. F. Harrison and  W.G.\ Scott,
Phys.\ Lett.\ {\bf B333}, 471 (1994);
P. F. Harrison, D. H. Perkins and W. G. Scott,
Phys.\ Lett.\ {\bf B349}, 137 (1995),
ibid.\ {\bf B374}, 111 (1996)
and {\bf B396}, 186 (1997).

\bibitem{LB} For a short description of some of these experiments, see
{\em Neutrino '96}, {\em op.\ cit.}

\bibitem{yasuda} O. Yasuda, hep-ph/9804400.

\bibitem{teshima} T. Teshima and T. Sakai,
preprint CU-TP/98-05, hep-ph/9805386.
 
\bibitem{bilenky} S. M. Bilenky, C. Giunti and C. W. Kim,
Astropart.\ Phys.\ {\bf 4}, 241 (1996), hep-ph/9505301.

\bibitem{ref:CHOOZ}
M. Apollonio et al.\ (CHOOZ Collaboration), 
Phys.\ Lett.\ {\bf B 420}, 397 (1998). 
 
\bibitem{bugey}
B. Achkar et al., Nucl.\ Phys.\ {\bf B 434}, 503 (1995).

\bibitem{K2K}
Y. Oyama, talk at YITP Workshop on Flavour Physics, 
Kyoto, January 1998, hep-ex/9803014.

\end{thebibliography}
\end{document}